\documentclass[prd,twocolumn,superscriptaddress,showpacs,floatfix,preprintnumbers,nofootinbib]{revtex4}

\usepackage{graphicx,amsmath,amssymb,bm,multirow}
\usepackage{psfrag}

\newcommand{\beq}{\begin{equation}}
\newcommand{\eeq}{\end{equation}}
\newcommand{\bea}{\begin{eqnarray}}
\newcommand{\eea}{\end{eqnarray}}
\newcommand{\e}{{\rm e}}
\newcommand{\jb}[2]{{\rm j}_{#1}(#2)}
\newcommand{\hone}[2]{{\rm h}^{(1)}_{#1}(#2)}

\newcommand{\Jb}[2]{{\rm J}_{#1}(#2)}
\newcommand{\Hone}[2]{{\rm H}^{(1)}_{#1}(#2)}

\newcommand{\veps}{\varepsilon}
\newcommand{\half}{{\textstyle\frac{1}{2}}}
\newcommand{\mat}[4]{ \left(
                          \begin{matrix}#1 & #2 \cr     %
                                  #3 &  #4 \end{matrix}
                          \right)   }
\newcommand{\threej}[6]{ \left(
                          \begin{matrix}#1 & #2 & #3 \cr     
                                  #4 &\! #5 \!& #6 \end{matrix}
                          \right)   }
\newcommand{\nn}{\nonumber}

\begin{document}
\preprint{FZJ-IKP-TH-2005-28}

\title{Scalar Casimir effect between Dirichlet spheres 
        or a plate and a sphere  }

\author{Aurel Bulgac} 
\affiliation{Department of Physics, University of Washington,
Seattle, WA 98195-1560, USA}
\author{Piotr Magierski}
\affiliation{Department of Physics, University of Washington,
Seattle, WA 98195-1560, USA}
\affiliation{Faculty of Physics, Warsaw University of Technology,
            ul. Koszykowa 75, 00-662 Warsaw, Poland}

\author{Andreas Wirzba}
\affiliation{Institut f\"ur Kernphysik (Theorie),
               Forschungszentrum J\"ulich, D-52425 J\"ulich,
               Germany }

\date{\today }

\begin{abstract}
  We present a simple formalism for the evaluation of the Casimir energy for
  two spheres and a sphere and a plane, in case of a scalar fluctuating field,
  valid at any separations. We compare the exact results with various
  approximation schemes and establish when such schemes become useful. The
  formalism can be easily extended to any number of spheres and/or planes
  in three or arbitrary dimensions,
  with a variety of boundary conditions or nonoverlapping
  potentials/nonideal reflectors.
\end{abstract}


\maketitle  


\section{Introduction}
In 1948 the Dutch physicist Hendrik~Casimir predicted 
the existence of a very peculiar effect, the
attraction between two metallic uncharged parallel plates in 
vacuum~\cite{casimir}. The existence of such an attraction has been confirmed
experimentally {\em with high accuracy} 
only recently \cite{lamoureaux,bordag}. However, nearly all modern 
experiments 
(the noted exceptions are the two-cylinder work of Ref.~\cite{ederth}
and the two-plate experiment of Ref.~\cite{bressi})
study the attraction  between a metallic sphere and a metallic plate which 
are much simpler to align
than two plates, but much harder to calculate. In fact, with the exception of
the proximity-force {\em approximation}~\cite{prox1,prox2}, which is only 
applicable
for vanishing separation, there does not exist a theoretical prediction
for the Casimir energy  of the sphere-plate system 
as function of the distance.

The origin of this attractive force can be traced back to the modification
in the spectrum of zero point fluctuations of the electromagnetic
field when the separated mirrors are brought into close
distance. 
Similar phenomena are expected to exist for various other
(typically bosonic) fields \cite{others,kardar} and the corresponding
forces are referred to as Casimir or fluctuation-induced interactions. A
related interaction arises when the space is filled with
fermions, which is particularly relevant to the
physics of neutron stars \cite{nm,mh,earlier,bw01,bhmwy03,bmw05} 
and quark gluon plasma~\cite{qgp}.  

Since the Casimir effect between a sphere and a plate
is the experimentally interesting
but theoretically difficult case, because of the electromagnetic nature of
the fluctuating fields, the corresponding Casimir effect 
for a real scalar field between two spheres or one sphere and a plate 
came into the focus of 
theoretical 
research~\cite{sspra,ssprl,gies1,gies2,emig,jaffe,scardI,scardII,scardIII}.
Therefore 
we will here focus on the exact and semiclassical calculation of this scalar 
Casimir effect 
between spheres or spheres and plates 
in three  spatial dimensions, where we assume
Dirichlet boundary conditions on the spheres and plates.
This scenario can
be trivially extended to the case of Neumann or other boundary conditions,
to two-dimensional systems of disks and/or lines and
to analogous systems
in arbitrary dimensions. 
We will show that the Casimir force calculation for the two-sphere case 
incorporates the sphere-plate geometry 
as special case. Note that, with the
exception of the
numerical results of Refs.\,\cite{gies1,gies2} (see also \cite{emig}), 
which do not yet extend to large separations, no exact result
exists in the literature for 
the Casimir effect for a real scalar field under Dirichlet boundary conditions,
neither between a sphere and a plate nor between two spheres. 
For small separations between the sphere and the plate, the above-mentioned
proximity-force approximation \cite{prox1,prox2} 
can  be applied. Its justification at small separations 
has been provided by many authors
using various theoretical techniques. For instance,
in Refs.\,\cite{sspra,ssprl} semiclassical methods in the framework of
the {\em Gutzwiller trace formula}~\cite{gutbook} have been used, 
in Ref.\,\cite{balian3} the proximity-force approximation
for the electromagnetic case
has been derived from the {\em multiple scattering expansion} of
Ref.\,\cite{balian12},
in Refs.\,\cite{gies1,gies2} the {\em world-line approach} in the
framework of the Feynman path integral has been applied, 
and in 
Refs.\,\cite{jaffe,scardI,scardII,scardIII} 
a ray-dynamical approach in terms of {\em optical paths} ({\it i.e.} 
closed but not
necessarily periodic orbits) has been employed.

Here, we will present an evaluation of the 
scalar Casimir problem that is based on 
quantum mechanics, without any semiclassical approximation made beforehand.
It utilizes
the Krein formula \cite{krein,uhlenbeck} 
as a bridge between the spectral density on the one hand
and the problem of  scattering of a point particle between spheres
in three dimensions (or disks in two 
dimensions)  on the other hand.
It has to be emphasized that in this case the Casimir calculation is not
plagued 
by the removal of diverging ultra-violet contributions.  This 
is related to the fact that the Krein formula is exact, 
and that the determinants of the
S-matrix and the corresponding inverse 
multi-scattering matrix for a system of $N$
nonoverlapping spheres (or disks), which both are manifestly known, 
are finite
\cite{wreport,wh98,hwg97}.
In this work we do not consider the material dependent stress of,
 {\it e.g.},  the deformation of a  single  
 spherical shell which was discussed in Ref.\,\cite{graham}.

The paper is organized as follows: first, we formulate the problem in terms of
the density of states of the scalar field and relate it to the S-matrix of the
system using the Krein formula. In Section \ref{sec:twosphere} 
we focus on the particular
realization of the Casimir effect for the sphere-sphere and the sphere-plate
system. Section \ref{sec:large-distance} 
is devoted to investigations of the large-distance limit
between spheres where the asymptotic expressions are derived and discussed.
Section \ref{sec:semiclassical} 
discusses the link between the presented approach and the
semiclassical methods based on the Gutzwiller trace formula. Finally, the
numerical results, approximate expressions and conclusions are presented in
Sections \ref{sec:results} and \ref{sec:conclusions}. 
For the sake of completeness, the derivation  of 
proximity-force
approximation for the sphere-sphere and sphere-plate system is given in 
Appendix~\ref{app:PFA} and a 
comparison to the two-dimensional two-disk and
disk-line systems can be found in Appendix~\ref{app:twodim}.

\section{The modified Krein formula}
Our main goal is to reach a qualitative understanding of
the scalar Casimir energy at zero temperature in the case of more 
complicated geometries than the original
two-plate system.  
For that purpose let us consider
the fluctuating real scalar field between
$N$ nonoverlapping, nontouching, 
impenetrable spheres of radii $a_i$  ($i=1,\dots,N$).
It is assumed that that the scalar field is noninteracting and
is subject to Dirichlet boundary conditions on the surfaces of the spheres.
The spheres are 
positioned at fixed relative distances 
$r_{ij}=L_{ij}+a_i+a_j >  a_i+a_j$ between their centers; $L_{ij}$ is then
the shortest relative distance between their surfaces, and $\vec{r}_{ij}$ is
the center-to-center distance vector, which includes also the information
about
the spatial orientation.
In order to
calculate the Casimir energy we shall represent the 
smoothed bosonic density of states of the scalar field
as a function of the energy $\veps$ 
(smoothing is over an energy
interval $\Delta\veps$ larger than the level spacing 
in the big volume $V$ of the
entire system):
\bea
  g\bigl(\varepsilon,\{a_i\},\{\vec{r}_{ij}\}\bigr)=&&
  g_0(\varepsilon)+\sum_{i=1}^N g_W(\varepsilon,a_i) \nn\\
&&\text{}+ g_{\rm C}\left(\varepsilon, \{a_i\},\{\vec{r}_{ij}\}\right),
  \label{dos}
\eea
where $g\bigl(\varepsilon,\{a_i\},\{\vec{r}_{ij}\}\bigr)$
is the total  density of states of the scalar
field,
$g_0(\varepsilon)$ is the density of states in the absence of all
scatterers, and $g_W(\varepsilon ,a_i)$ is the correction
to the density of states arising from the presence of one sphere (sphere $i$).
Clearly $\sum_{i=1}^N g_W(\varepsilon,a_i)$ is the correction 
due to the $N$ spheres 
infinitely far apart from each other 
that sums up  the excluded volume effects, surface
contributions and Friedel oscillations caused by
each of the obstacles separately. 
 Finally 
$g_{\rm C}\bigl(\veps, \{a_i\},\{\vec{r}_{ij}\})\bigr)$ is the remaining
part, which is of central interest to us here.
It vanishes in the limit of infinitely separated scatterers and
is the only term in the density of states which reflects
the relative geometry-dependence of the problem. Only this term contributes
to the Casimir energy. 

Strictly speaking, only the smoothed 
level densities $g_W(\veps)$ and $g_{\rm C}(\veps)$ are 
finite, whereas the level densities $g(\veps)$ and
$g_{0}(\veps )$ are infinite, as they are proportional to the volume
$V$ of the entire space. This redundant divergence can be handled
easily by considering the smoothed quantities first in a very big box,  
the volume of which is
subsequently taken to infinity\cite{wreport}.

Now we will use the Krein formula \cite{krein,uhlenbeck} which 
provides a link between the ($N$-body) scattering matrix $S_N(\veps )$ of
a point-particle scattering off $N$ spheres 
and the change in the 
density of states due to the presence of $N$
scatterers, namely
\bea
  &&\delta g\bigl(\veps,\{a_i\},\{\vec{r}_{ij}\}\bigr) = 
  g\bigl(\veps,\{a_i\},\{\vec{r}_{ij}\}\bigr) -g_0(\veps) \nn \\
  &&= \frac{1}{2\pi {\rm i}} \;\frac{d \ }{d \veps }
  \ln \det S_N\bigl(\veps,\{a_i\},\{\vec{r}_{ij}\}\bigr)\,.
  \label{Krein}
\eea
Note that $\ln \det S_N\bigl(\veps,\{a_i\},\{\vec{r}_{ij}\}\bigr)/2 {\rm i}$ is
nothing else than the total phase shift of the scattering problem. 
The geometry-dependent
part of the density of states can now be extracted from the genuine
multi-scattering determinant. In this way the calculation  is mapped
onto a quantum mechanical {\em billiard} problem that classically corresponds 
to the  hyperbolic (or even chaotic) point-particle scattering off
$N$ spheres~\cite{hwg97} (or $N$ disks in two 
dimensions
~\cite{Eck_org,gasp,Cvi_Eck_89,pinball,aw_chaos,aw_nucl,vwr94,rvw96}).

As shown in Refs.\,\cite{wreport,wh98,hwg97}, 
the determinant of the $N$-scatterer S-matrix,
$S_N\bigl(\veps,\{a_i\},\{\vec{r}_{ij}\}\bigr)$, 
factorizes into the product of the determinants of the single-scatterer 
S-matrices and
the ratio of the determinants of the inverse multi-scattering 
matrix~\cite{wreport,wh98,hwg97}  of Korringa--Kohn--Rostoker (KKR) 
type~\cite{kkr,Lloyd,Lloyd_smith,berry81}  
$M(k)=M(k,\{a_i\},\{\vec{r}_{ij}\})$ 
in the complex wave
number ($k=|\vec{k}|$) plane:
\beq
  \det S_N\bigl(\veps,\{a_i\},\{\vec{r}_{ij}\}\bigr) 
  =
\left\{\prod_{i=1}^N \det S_1(\veps,a_i) \right\}
   \frac{\det M(k^*)^\dagger}
     {\det M(k) }.
  \label{detS}
\eeq
The formula (\ref{detS}) 
holds in the case when the scattering modes are free massless fields as
well as in the case of
free nonrelativistic fields with a mass $m$. 
Both cases imply different energy dispersion relations, 
$\veps=\hbar\omega=\hbar c  k$ in the massless case or
$\veps=\hbar^2k^2/(2m)$  in the nonrelativistic scenario, respectively.

Although the 
involved matrices  are infinite-dimensional, all
determinants 
are well-defined, as long as the number of spheres is finite and the spheres
do neither overlap nor touch.  This  follows from the
trace-class property~\cite{rs4,bs_adv} of the matrices  
$S_N-\openone$, $S_i-{\openone}$ 
and $M-{\openone}$ which was  shown in 
Refs.~\cite{wreport,wh98,hwg97}.~\footnote{Trace-class operators (or matrices)
are those, in general, non-Hermitian operators (matrices) of a separable
Hilbert space which have an absolutely convergent trace in every orthonormal
basis. Especially the determinant $\det (\openone + z {A})$ exists 
and is  an
entire function of $z$, if ${A}$ is trace-class.}

Inserting the exact expression (\ref{detS}) into the original Krein
formula (\ref{Krein}), using the decomposition (\ref{dos}) 
and identifying the ``Weyl-type'' density of states
with the phase shift of the corresponding single scatterer 
\beq
   g_W(\veps,a_i) = \frac{1}{2\pi {\rm i}} 
  \;\frac{{\rm d} \ }{{\rm d} \veps } \ln \det S_1(\veps,a_i)\,,
\eeq
one finds a new Krein-type exact 
formula~\cite{bw01} which directly links the geometry-dependent
part of the density of states with the inverse multi-scattering matrix
\[
  g_{\rm C}\bigl(\veps,\{a_i\},\{\vec{r}_{ij}\}\bigr) =
  \frac{{\rm d} \ }{{\rm d} \veps }  
  \frac{-1}{\pi}{\rm Im}
  \ln \det M\bigl(k(\veps),\{a_i\},\{\vec{r}_{ij}\}\bigr)\nn\\
\]
or 
\beq
  {\cal N}_{\rm C}\bigl(\veps,\{a_i\},\{\vec{r}_{ij}\}\bigr) 
  =-\frac{1}{\pi}{\rm Im}
  \ln \det M\bigl(k(\veps),\{a_i\},\{\vec{r}_{ij}\}\bigr)
  \label{modKreinNc}
\eeq
for the integrated geometry-dependent part of the density of states
\beq
  {\cal N}_{\rm C}\bigl(\veps,\{a_i\},\{\vec{r}_{ij}\}\bigr)
  =\int_0^\veps {\rm d}
  \veps' g_{\rm C}\bigl(\veps',\{a_i\},\{\vec{r}_{ij}\}\bigr) \,.
  \label{KreinNc}
\eeq
The exact formula (\ref{modKreinNc}) 
is the central expression of this paper. The Casimir energy
itself follows via the integral
\bea
  {\cal E}_{\rm C} &=& \int_0^\infty {\rm d}\veps\, \half\veps\,  
  g_{\rm C}(\veps) 
  =-\half\int_0^\infty {\rm d}\veps\, 
  {\cal N}_{\rm C}(\veps) \nn\\
  &=&\frac{1}{2\pi} \int_0^\infty {\rm d}\veps\, 
  {\rm Im}\,\ln\det M\!\left(k(\veps)\right)\nn\\
  &=&\frac{1}{4\pi{\rm i}} \Biggl[\int_0^{\infty(1+{\rm i}0_+)} {\rm d}\veps\, 
  \ln \det M\!\left(k(\veps)\right)\nn\\
  &&\qquad\text{} -\int_0^{\infty(1-{\rm i}0_+)} {\rm d}\veps\, 
  \ln\det M\!\left(k^\ast(\veps)\right)^\dagger\Biggr]\,.
  \label{Ecint}
\eea
In the massless case $\veps=\hbar c k$, this expression can be Wick-rotated
({\it i.e.}, $k\to {\rm i}\,k_4$ for the first term and 
$k\to -{\rm i}\,k_4$ for the second term of the last relation)
to give~\footnote{The Wick-rotations are allowed, since 
 $\det M(k)$ has poles in the lower complex
$k$-plane only, whereas $\det M(k^\ast)^\dagger$ has poles in the 
upper half-plane~\cite{wreport,wh98,hwg97}. 
Furthermore, the integrals over
the circular arcs vanish,
since $\ln\det M(k)$  and $\ln\det M(k^\ast)^\dagger$ are
exponentially suppressed for ${\rm Im }\,k\to\pm \infty$, respectively; 
see, {\it e.g.}, the semiclassical expression
(\ref{GutzVoros}).}
\bea
  {\cal E}_{\rm C} &=& 
  \frac{\hbar c }{2\pi}
  \int_0^\infty {\rm d} k_4\, \ln\det M({\rm i}k_4)  \,,
  \label{EcWick}
\eea
since $\det M(k)=\det M((-k)^\ast)^\dagger$ and therefore
$\det M({\rm i}k_4)= \det M({\rm i}k_4)^\dagger$ if $k_4$ 
real.~\footnote{For the same reason, one can show the corollary that
all the corresponding integrals over odd powers of $k$ have 
to vanish:
\begin{eqnarray*}
  0&=&
  \frac{\hbar c }{2\pi}
  \int_0^\infty {\rm d} k\, k^{2n+1} {\rm Im}\,\ln\det M(k)\\
  & =& {\rm i}(-1)^n\frac{\hbar c }{4\pi}
  \int_0^\infty \!\!{\rm d} k_4\, k_4^{2n+1}\left[
\ln\det M({\rm i}k_4) -\ln\det M({\rm i}k_4)^\dagger\right].
\end{eqnarray*}
Thus, the Casimir energy over modes with a nonrelativistic dispersion
$\veps=\hbar^2 k^2/(2m)$ integrates to zero, unless there exists a finite upper
integration limit, as {\it e.g.} the Fermi momentum in the fermionic Casimir
effect studied in Ref.~\cite{bw01,bmw05}.
}

Using the explicit formulas for the KKR--type matrix from
Ref. \cite{wreport,wh98,hwg97}, one can compute numerically ${\cal E}_{\rm C}$
for various arrangements of hard spherical (or circular in 2D) scatterers. 
For a point-particle scattering off 
$N$ nonoverlapping nontouching spheres (under Dirichlet boundary conditions)
\bea
  && M^{jj'}_{lm,l'm'} 
  = \delta^{jj'}\delta_{ll'}\delta_{mm'}
                 + (1-\delta^{jj'})\, \sqrt{4\pi}\,{\rm i}^{2m+l'-l}
  \nn\\
  &&\text{}\times
  \sqrt{(2l\!+\!1)(2l'\!+\!1)}\,
  \left({\textstyle\frac{a_j}{a_{j'}}} \right)^2
  \frac{ \jb{l}{k a_j} }{\hone{l'}{k a_{j'}}}\nn\\
  &&\text{}\times 
  \sum_{l'' =0}^{\infty}\sum_{m''=-l'}^{l'}
  {\rm D}^{l'}_{m'\!,m''}(j,j')\,
  \hone{l''}{k r_{jj'}}\,
  {{\rm Y}_{l''}^{m\!-\!m''}\!\bigl( \hat r^{(j)}_{jj'}\bigr)}
  \nn\\
  &&\text{}\times 
  \sqrt{2l''\!+\!1}
  \,{\rm i}^{l''}
  \threej{l''}{l'}{l}{0}{0}{0}
  \threej{l''}{l'}{l}{m\!-\!m''}{m''}{-m}
 \label{M_N-sphere}
\eea
is the inverse multi-scattering matrix~\cite{hwg97}.
Here $j,j'=1,2,\dots,N$ are the labels of the $N$ spheres,
$l,l',l''=0,1,2,\dots$ are the angular momentum quantum numbers, and 
$m,m',m''$ the pertinent magnetic quantum numbers. 
${\rm D}^{l'}_{m'\!,m''}(j,j')$
is a Wigner rotation matrix which transforms the local coordinate system from
sphere $j'$ to the one of sphere $j$, $\jb{l}{kr}$ and $\hone{l}{kr}$ 
are the spherical Bessel and Hankel functions of first kind 
respectively, 
${\rm Y}_{l''}^{m\!-\!m''}\!\bigl( \hat r^{(j)}_{jj'}\bigr)$ is 
a  spherical harmonic (where $\hat r^{(j)}_{jj'}$ is the
unit vector, measured in the local frame of sphere $j$, pointing
from sphere $j$ to sphere $j'$), and the 
3j-symbols~\cite{landolt}
result from the angular momentum coupling.

By definition, 
the inverse multi-scattering matrix incorporates the pruning rule that two 
successive scatterings have to take place at different scatterers (see
the $(1-\delta^{jj'})$ term in Eq.\,(\ref{M_N-sphere})). This alternating
pattern between a single-scatterer T-matrix and the successive 
propagation to a new scatterer~\cite{wreport,Lloyd,Lloyd_smith} distinguishes
the KKR-type method from the multiple scattering 
expansion of Refs.\,\cite{balian12,balian3}.

\section{The two-sphere and the sphere-plate problem}
 \label{sec:twosphere}
\begin{figure*}[h,t,b]

\centerline{\includegraphics[width=8cm]{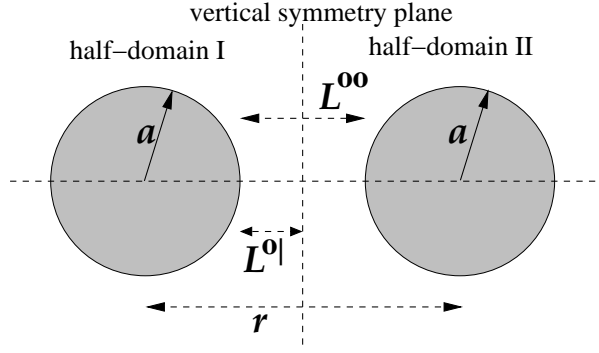}}

\caption{ Two identical spheres of radius $a$ at a center-to-center separation
$r$. The vertical symmetry plane, the two half-domains and the
surface-surface separation $L^{{\rm oo}}$ in the global domain and 
$L^{{\rm o|}}$ in one of the half-domains are shown.}

\label{fig:twosphere}

\end{figure*}

In the case of two spheres the system possesses 
a continuous symmetry, 
{\it i.e.}
$C_{\infty\, v}$ in  crystallography group theory notation~\cite{hamermesh},
associated with rotations with respect to the axis joining the centers
of the spheres (and an additional reflection symmetry 
with respect to any plane containing
this symmetry axis).
As a consequence the KKR-matrix is separable 
with respect to the magnetic quantum number $m$ (in fact, it 
depends on its modulus $|m|$ only)~\cite{hwg97}. 
In the global domain, it is given as
\beq
   \mat{{M}^{11}}{{M}^{12}} 
       {{M}^{21}}{{M}^{22}}_{lm, l' m'}
   \!\!\!\!\!=\delta_{m m'}
  \mat{ \delta_{l l'} }{ A_{ll'}^{12\, (m)} }
  { A_{ll'}^{21\, (m)}   }{ \delta_{l l'} }
 \label{M_2-sphere}
\eeq
with
\bea
  &&  A_{ll'}^{jj'\, (m)}= (-1)^{l'}\, {\rm i}^{l'-l}
  \frac{{a_j^2\, \jb{l}{k a_j}}}{{ a_{j'}^2 \hone{l'}{k a_{j'}}}}
  \sqrt{(2l\!+\!1)(2l'\!+\!1)}\nn\\
  &&\text{}\times\sum_{l''}{\rm i}^{l''}(2l''\!+\!1)
  \threej{l''}{l}{l'}{0}{0}{0}\threej{l''}{l}{l'}{0}{m}{-m}
  \hone{l''}{k r}.\nn\\
  \label{A-expression}
\eea
If the spheres have moreover the same radius $a_1=a_2\equiv a$, 
there exist also a two-fold
reflection symmetry with respect to the vertical symmetry plane.
This additional symmetry makes the  
total symmetry of the system to be  $D_{\infty\, h}$ 
in the crystallography group
notation~\cite{hamermesh}~\footnote{Note that Fig.\,\ref{fig:twosphere} is
rotated by 90 degree relative to the conventions of Ref.\,\cite{hamermesh},
such that our vertical symmetry plane is called ``horizontal'' there.},
which is a simply product of the $C_{\infty\, v}$ 
group and the inversion (and rotation by $\pi$) with respect to the
point of intersection between the symmetry axis 
and vertical symmetry plane.

Therefore the global domain of the two-sphere system
can be split into two half-domains, separated by this 
plane, see Fig.\,\ref{fig:twosphere}, and 
all the (scattering) wave functions can decomposed into symmetric
and antisymmetric ones with respect to the vertical symmetry plane. The
symmetric wave functions are subject to Neumann boundary conditions and
the antisymmetric are subject to Dirichlet 
boundary conditions on this symmetry plane. 
Thus there exist two KKR matrices in the 
the half-domain~\cite{hwg97}, one corresponding to the Neumann case
(N) and the other, with the additional minus sign, 
to the Dirichlet case (D):
\bea
  \left. M^{{\rm oo}\,(m)}_{ll'}\right|_{\rm N} &=& \delta_{ll'} 
  + A_{ll'}^{(m)}\,,
  \label{M_Neumann}
  \\
  \label{M_Dirichlet}
   \left. M^{{\rm oo}\,(m)}_{ll'}\right|_{\rm D} 
   &=& \delta_{ll'} - A_{ll'}^{(m)}\,,
\eea
where $A_{ll'}^{(m)}\equiv \left. A_{ll'}^{12\,(m)}\right|_{a_1=a_2\equiv a}
= \left. A_{ll'}^{21\,(m)}\right|_{a_1=a_2\equiv a}$.
Furthermore 
the KKR determinant $\det M^{{\rm oo}}(k,a,r)$ $\equiv$ 
$\det M^{{\rm o}_1{\rm o}_2}(k,a_1\!=\!a,a_2\!=\!a,r)$ 
of the full domain factorizes into the product of
the determinants of these Neumann and Dirichlet KKR matrices
\bea
  && \det M^{{\rm oo}}(k,a,r)=\nn\prod_{m=-\infty}^\infty
  \det M^{{\rm oo}\,(m)}(k,a,r) \\
  && = \prod_{m=-\infty}^\infty
  \det \left. M^{{\rm oo}\,(m)}(k,a,r)\right|_{\rm N}
                     \det \left. M^{{\rm oo}\,(m)}(k,a,r)\right|_{\rm D} \nn\\
  &&\equiv \det \left. M^{{\rm oo}}(k,a,r)\right|_{\rm N}
                     \det \left. M^{{\rm oo}}(k,a,r)\right|_{\rm D}\,\,.
\eea
Thus the two-sphere system contains the Dirichlet 
sphere-plate system as special case, namely in the symmetric limit
$a_1=a_2=a$, the sphere-plate system
is equal the Dirichlet case in the half-domain, and the pertinent
multi-scattering determinant $\det M^{{\rm o|}}$ of the sphere-plate system
is just given as
\beq
  \det M^{{\rm o|}}\Bigl(k,a,L^{\rm o|}\Bigr) =\left. \det M^{{\rm oo}}
  \Bigl(k,a,r\!=\!2(L^{\rm o|}\!+\!a)\Bigr)\right|_{\rm D}\,.
\eeq
Note that the shortest surface-to-surface distance in the symmetric 
two-sphere case is given by $L^{\rm oo}=r-2a $, whereas the shortest 
surface-to-surface distance in the sphere-plate case is
$L^{\rm o|}=\half r-a $, see Fig.\,\ref{fig:twosphere}.

The exact expressions for the Casimir energy of the two-sphere Dirichlet 
problem,
the symmetric two-sphere Dirichlet problem and the sphere-plate Dirichlet 
problem are given by the following integrals, respectively:
\bea
  &&{\cal E}_{\rm C}^{{\rm o}_1 {\rm o}_2}(a_1,a_2,L)\nn\\
  && = \frac{\hbar c }{2\pi}
 \int_0^\infty \!\!\!{\rm d} k_4 \ln\det M^{{\rm o}_1{\rm o}_2} 
  ({\rm i}k_4,a_1,a_2,r\!=\!L\!+\!a_1\!+\!a_2)\,, \nn \\
  &&{\cal E}_{\rm C}^{{\rm o o}}(a,L)= \frac{\hbar c }{2\pi}
  \int_0^\infty \!\!\!{\rm d} k_4 \ln\det M^{\rm o o} 
  ({\rm i}k_4,a,r\!=\!L\!+\!2a)\,,\nn \\
  &&{\cal E}_{\rm C}^{{\rm o |}}(a,L)\nn\nn\\&&
= \frac{\hbar c }{2\pi}
  \int_0^\infty \!\!{\rm d} k_4 \ln\det \left. M^{\rm o o} \bigl
  ({\rm i}k_4,a,r\!=\!2(L\!+\!a)\bigr)\right|_{\rm D}. \nn\\
  \label{E_sp_exact}
\eea
In practice, these expressions have to be numerically integrated 
up to an upper
value  $k_4^{\rm max}$ which should be chosen large enough, 
such that the numerical value
of the integral is stable for some specified range of decimal places.
For the sphere-plate case, $k_4^{\rm max}\sim 10/L$ specifies a good choice.
In order that the evaluation of the determinant of the
matrix $\left. M^{\rm oo}_{ll'}\bigl(m;k,a,r\!=\!2(L\!+\!a)
\bigr)\right|_{\rm D}$ 
is stable, the upper value of $k$ induces a maximal value $l_{\rm max}$ 
for $l,l'$ 
and $m$ (with $|m|\leq l,l'\leq l_{\rm max}$), namely \cite{berry81,wreport}
\beq
  l_{\rm max} \geq \frac{{\rm e}}{2} k_4^{\rm max} a \approx  14 a/L\,.
 \label{lmax}
\eeq
For small values of the separation $L$, the maximal angular momentum and
therefore the size of the KKR matrices (which scale with $l^2\times l$ where
the last factor results from the separable $m$ quantum number) becomes
rapidly very large and limits the range of applicability of this 
numerical computation of the exact integral to medium and large
values of $L$, say, $ L> 0.1 a$, chiefly because of the handling of 
the 3j-symbols which scale with $l^4$.

\section{Large-distance limit}
\label{sec:large-distance}
As shown in Ref.\,\cite{bw01},
it is possible to obtain significantly simpler expressions for the
(integrated) density of states in the limit of very large separation or very
small scatterers. If the wave length $\lambda =2\pi/k$ 
is much larger than the radii of the scatterers one can
show that the KKR--matrix $M(k)$ is given by 
\beq
  [ M (k)]^{jj'}  \approx
  \delta^{jj'} - (1-\delta^{jj'}) f_j(k)
  \frac{ \exp ({\rm i} kr_{jj'})}{r_{jj'}}   \label{eq:kkr0}
\eeq
(see \cite{rww96} for the analog in the 2D case),
where the indices $j,j'=1,\dots,N$ denote the scatterers, $r_{jj'}$ is
the distance between their centers and
$f_j(k)$ is the $s$-wave scattering amplitude on the
$j$--th scatterer.

In the case of two
identical spheres of radius  $a$, with their centers located at 
the distance $r$ 
apart, one obtains 
\bea
 \det M(k)&=&1 - \frac{a^2}{r^2}\, \exp\left[{\rm i} k(2 r -2a )\right]
  \label{twosphere_glob}\\
  &=&\left.\det M(k)\right|_{N} \left.\det M(k)\right|_{D}
  \nn\\
  &=&  \left\{ 1 + \frac{a}{r}{\rm e}^{{\rm i} k(r -a )}\right\}
  \left\{ 1 - \frac{a}{r} {\rm e}^{{\rm i} k(r -a )}\right\}.
  \label{twosphere_half}
\eea
As usual, the
determinant of this two-sphere system in the global domain factorizes
into two sub-determinants calculated for 
one of the half-domains, one subject to Neumann boundary conditions and the
other (with the minus sign) subject to Dirichlet boundary conditions.

As mentioned, this expression can be derived from Eqs.\,
(\ref{M_Neumann}-\ref{M_Dirichlet}) 
in the case of large center-to-center separations of spheres, $k r \gg 1$: 
we can make use of 
the asymptotic expression for the spherical
Hankel function~\cite{Abramowitz}
\beq
  \hone{l}{kr}\sim \frac{\exp({\rm i}kr)}{{\rm i}^{l+1}kr}\,,
  \label{Hankel_asymp}
\eeq
which is actually exact for $\hone{0}{kr}$,
such that the inverse multi-scattering matrix becomes
\bea
  && M^{{\rm oo}\,(m)}_{l_1 l_2} (k, a, r) 
  \sim \delta _{l_1 l_2} \pm
  \frac{\jb{l_1}{ka}}{\hone{l_2}{ka}}\, 
  \frac{\sqrt{(2l_1\!+\!1)(2l_2\!+\!1)}}{{\rm i}^{l_1+l_2}}\nn\\
  &&\times\frac{\exp ({\rm i}kr)}{{\rm i} kr}\sum_{l=0}^\infty (2l\!+\!1)
  \threej{l_1}{l_2}{l}{m}{-m}{0}
  \threej{l_1}{l_2}{l}{0}{0}{0} .
  \label{M-expand}
\eea
Using now the orthogonality relation for the $3j$--symbols~\cite{landolt}
\[
   \sum_{l=0}^\infty (2l\!+\!1)\! \threej{l_1}{l_2}{l}{m_1}{m_2}{m}
               \!\threej{l_1}{l_2}{l}{m_1'}{m_2'}{m}\!\!
               =\delta_{m_1 m_1'} \delta_{m_2 m_2'},
\]
we get the asymptotic result
\bea
   &&M^{{\rm oo}\,(m)}_{l_1 l_2} (k, a, r) 
   \sim \delta _{l_1 l_2} \pm
  \frac{\jb{l_1}{ka}}{\hone{l_2}{ka}}\nn\\
  &&\qquad\text{}\times
  \frac{\sqrt{(2l_1+1)(2l_2+1)}}{{\rm i}^{l_1+l_2}}\,
  \frac{\exp ({\rm i}kr)}{{\rm i}kr}\, \delta _{m0} .
 \label{M_asymp_res}
\eea
Since for $kr\gg 1$ the only  
nontrivial  matrix $M^{{\rm oo}\,(0)}_{l_1 l_2}$ 
is separable in $l_1$ and $l_2$, 
the corresponding  determinant is simply given by
\bea
  &&\det M^{\rm oo}(k ,a,r)\nn\\
  &&\sim 1 \!-\! \frac{\exp (2{\rm i}kr)}{({\rm i}kr)^2}
  \left [ \sum _{l=0}^\infty (-1)^l(2l\!+\!1)
  \frac{\jb{l}{ka}}{\hone{l}{ka}}
  \right ]^2\nn\\
  &&\equiv 1 - \frac{\exp (2{\rm i}kr)}{(kr)^2}\, \left[-{\rm i}A(ka)\right]^2.
  \label{detM_asymp_glob}
\eea
for the identical two-sphere case and by
\bea
  &&\det M^{\rm o|}(k ,a,L) \equiv 
  \det \left.M^{\rm oo}\bigl(k ,a,r\!=\!2(L\!+\!a)\bigr)\right|_{\rm D}
  \nn \\
  &&\sim 1 \!-\! \frac{\exp ({\rm i}kr)}{{\rm i}kr}
  \sum _{l=0}^\infty (-1)^l(2l\!+\!1)
  \frac{\jb{l}{ka}}{\hone{l}{ka}} \nn\\
  &&\equiv 1 - \frac{\exp ({\rm i}kr)}{ kr}\, \left(-{\rm i}A(ka)\right)
  \label{detM_asymp_half}
\eea
for the sphere-plate case.
Here
\beq
  A(ka)\equiv  \sum _{l=0}^\infty (-1)^l(2l\!+\!1)
  \frac{\jb{l}{ka}}{\hone{l}{ka}}
  \label{Amplitude}
\eeq
is the multipole expansion of the scattering amplitude. 
For $ka \ll 1$, 
the ratio $\jb{l}{ka}/\hone{l}{ka}$ becomes~\cite{Abramowitz} 
\bea
   \frac{\jb{l}{ka}}{\hone{l}{ka}} &=& 
   \frac{{\rm i} (ka)^{ 2l+1} {\rm e}^{-{\rm i} ka}}
                             { 1^2 \times 3^2 \times \cdots\times 
            (2l-1)^2
                               \times (2l+1)}\nonumber\\    
        &&\text{}     + {\cal O}\left( (ka)^{2l+2+\delta_{l,0}} \right).
\eea
Thus the dominant effect comes from the $s$-wave scattering only and the
$l>0$ terms can be neglected.
Consequently, the scattering amplitude becomes
\bea
  -{\rm i}A(ka) &=& -{\rm i} \frac{\jb{0}{ka}}{\hone{0}{ka}} 
  +{\cal O}\left((ka^3)\right)\nn\\ 
  &=&  
  ka \exp(-{\rm i}ka)+{\cal O}\left((ka^3)\right) \,,
\eea
which implies
that (\ref{detM_asymp_glob}) becomes (\ref{twosphere_glob}) and
 (\ref{detM_asymp_half}) becomes the Dirichlet part of (\ref{twosphere_half}).

If the determinant in the global domain (\ref{twosphere_glob}) is inserted 
into the modified Krein formula (\ref{modKreinNc})
we get the following result for the integrated geometry-dependent part of
the density of states in the $s$-wave limit of the two-sphere case~\cite{bw01} 
(note
$r=L^{\rm oo}+2a$):
\beq
  {\cal{N}}_{s-{\rm wave}}^{\rm oo}(\veps )= 
  \frac{a^2}{\pi (L^{\rm oo}\!+\!2a)^2} 
  \sin \left[2 k(L^{\rm oo}\!+\!a)\right] 
  + {\cal{O}}\left((ka)^3\right) 
  \label{Nc_S_sphere} .
\eeq
Analogously one gets the $s$-wave limit for the Dirichlet sphere-plate system
by inserting the Dirichlet determinant of the half-domain, namely the 
second term of Eq.(\ref{twosphere_half}),  into  
(\ref{modKreinNc}):
\beq
  {\cal{N}}_{s-{\rm wave}}^{\rm o|} (\veps )= 
  \frac{a}{2 \pi(L^{\rm o|}\!+\!a)} \sin \left[k(2 L^{\rm o|}\!+\!a)\right] 
  + {\cal{O}}\left((ka)^3\right) 
  \label{Nc_S_plate} 
\eeq
[note $r=2(L^{\rm o|}+a)$].

Now, the Casimir energy for two identical Dirichlet spheres 
in the large $L=L^{\rm oo}$
limit follows simply by inserting ${\cal N}_{s-{\rm wave}}^{\rm oo}$, 
Eq.\,(\ref{Nc_S_sphere}), into the 
integral (\ref{Ecint}) and performing
the Wick-rotation as in (\ref{EcWick}):
\bea
  &&{\cal{E}}_{s-{\rm wave}}^{\rm oo}(a,L)
  =   -\hbar c \frac{a^2}{4\pi (L+a)(L+2a)^2}
  \nn\\
  &&= \kappa \frac{\pi a^2}{L^3} \left(\frac{90}{\pi^4}\right) 
  \frac{4}{\left(1+\frac{a}{L}\right)\left(1+\frac{2a}{L}\right)^2}\,,
\eea
where
\beq
  \kappa = -\frac{\hbar c}{16\pi^2}\,\frac{\pi^4}{90} =
  -\frac{\hbar c \pi^2}{1440}
\eeq
is the  prefactor of the Casimir energy 
\beq
  {\cal E}^{||}(L)=-\frac{\hbar c\pi^2}{2\times 720}\,\frac{A}{L^3}
\eeq
of the corresponding  scalar (Dirichlet) two-plate system where $A$ is
the area of the plates. 
Instead of performing the Wick rotation, 
one can compute these integrals along the real axis too. 
In this case one would have to include a convergence factor 
$\exp( -\eta\veps)$ and take the limit $\eta\rightarrow +0$
at the end of the calculation (in a similar manner to the 
Feynman prescription for propagators).

Similarly, the Casimir energy for the Dirichlet sphere-plate case 
in the large $L=L^{\rm o|}$
limit follows by inserting ${\cal N}_{s-{\rm wave}}^{\rm o|}$, 
Eq.\,(\ref{Nc_S_plate}), into the integral (\ref{Ecint}) and performing
the Wick-rotation:
\bea
 && {\cal{E}}_{s-{\rm wave}}^{\rm o|}(a,L)
  =   -\hbar c \frac{a}{4\pi (L+a)(2L+a)}
  \nn\\
  &&= \kappa \frac{\pi a}{L^2} \left(\frac{90}{\pi^4}\right) 
  \frac{2}{\left(1+\frac{a}{L}\right)\left(1+\frac{a}{2L}\right)}\,.
  \label{E_sp_s-wave}
\eea
The large-distance scaling is therefore proportional to $a/L^2$, in contrast to
the Casimir-Polder energy between a molecule and a conducting plane which
scales like $a^3/L^4$~\cite{polder}.
 The difference is associated with the fact 
that the Casimir-Polder energy
results from the induced dipole moment, whereas in the scalar scattering
the monopole term gives the dominant contribution.
In fact, if one omits the $s$-wave scattering
term and starts instead with the the $p$-wave term in the scattering function
$A(ka)$, the scalar Casimir energy for the sphere-plate system would show
a large-$L$ behaviour
\bea
  {\cal{E}}_{p-{\rm wave}}^{\rm o|}(a,L)
  &=& \kappa \frac{\pi a^3}{L^4} \left(\frac{90}{\pi^4}\right) 
  \frac{1}{\left(1\!+\!\frac{a}{L}\right)\left(1\!+\!\frac{a}{2L}\right)^2}\,
\eea
which is compatible with the $a^3/L^4$ scaling of the Casimir-Polder energy.
Thus the correct  large-distance 
behaviour of the scalar Casimir energy has nothing to do
with missing diffraction contributions 
(see Refs.~\cite{aw_chaos,aw_nucl,vwr94,rvw96}) to the semiclassical trace
formula 
as conjectured in Ref.~\cite{sspra} and repeated in 
Ref.~\cite{scardI}. It is rather based on the 
replacement of the semiclassical summation, which we will discuss in
the next section and which is valid when many partial amplitudes contribute, 
by the leading term(s) in the multipole expansion~\cite{rww96,bw01}.

In summary, the asymptotic expressions for the Casimir energy are given by
\beq
  {\cal{E}}_{\rm C}^{\rm oo}  
   ({L\gg a})\ \sim \ 4\times \frac{90}{\pi^4}\, 
  \frac{\kappa \pi a^2}{L^3} \approx  3.6958 \frac{\kappa\pi a^2}{L^3}
\eeq
in the symmetric two-sphere scalar Dirichlet case 
and by
\beq
  {\cal{E}}_{\rm C}^{\rm o|} ({L\gg a})\ \sim \  
  2\times \frac{90}{\pi^4}\, 
  \frac{\kappa\pi a}{L^2} \approx  1.8479 \frac{\kappa\pi a }{L^2}
  \label{E_sp_asymp}
\eeq
in the sphere-plate scalar Dirichlet case.

\section{The semiclassical approximation}
\label{sec:semiclassical}
The semiclassical approximation of the scalar two-sphere problem in
the framework of the Gutzwiller trace formula~\cite{gutbook} 
was pioneered 
in Refs.\,\cite{sspra,ssprl}. Here we will focus on the link between the
scattering approach and these semiclassical methods. 

The sphere-plate system at surface-to-surface separation $L$ 
is a special case of the sphere-sphere case
for two spheres  of radii $a_1$ and $a_2$ 
at center-to-center separation $R=L+a_1+a_2$
in the limit $a_2\to\infty$.
As shown in Ref.\,\cite{bw01}, the integrated density of states for the
two-sphere system follows
semiclassically from the Gutzwiller trace formula~\cite{gutbook}
\bea
  && {\cal N}^{{\rm o}_1 {\rm o}_2}_{\rm sc}(a_1,a_2,L,k)\nn\\
  &&=  \frac{1}{\pi} {\rm Im}\,
  \sum_{n=1}^{\infty}\frac{\exp(n{\rm i} 2 k L)}{n}
  \frac{1}{|\det \left(\left[{\sf M}(a_1,a_2,L)\right]^n
  -{\openone}\right) |^{1/2}} \nn \\
  &&= \frac{1}{\pi} {\rm Im}\,
  \sum_{n=1}^{\infty}\frac{1}{n}
  \frac{\exp(n{\rm i} 2 k L)}
  {|\Lambda_+(a_1,a_2,L)^n+\Lambda_-(a_1,a_2,L)^n-2|}\nn\\ 
  &&= \sum_{n=1}^{\infty}\frac{1}{n \pi} 
  \frac{\sin(n 2 k L)}{|\Lambda_+(a_1,a_2,L)^n+\Lambda_-(a_1,a_2,L)^n-2|} 
  ,
    \label{Gutztrace}
\eea
where the periodic orbit is the bouncing-orbit between the spheres and the
summation is over the repeats of this orbit. The matrix ${\sf M} (a_1,a_2,L)$ 
is the monodromy matrix of the sphere-sphere system and
$\Lambda_+(a_1,a_2,L)$ = $1/\Lambda_-(a_1,a_2,L)$ is the double--degenerate 
leading 
eigenvalue of this matrix, {\em i.e.}:
\bea
  \Lambda_{\pm}(a_1,a_2,L) &=& 1 + 
  2 L\left(\frac{1}{a_1}\!+\!\frac{1}{a_2}\right) 
  + 2\frac{L^2}{a_1 a_2}\nn\\
  &&\pm 
  \sqrt{\left(1\!+\!2 L\left(\frac{1}{a_1}\!+\!\frac{1}{a_2}\right)\! 
  +\!2\frac{L^2}{a_1 a_2}\right)^2-1}\,, \nn\\
 \label{Lambda_pm}
\eea
which is identical with Eq.\,(3.11) of Ref.\,\cite{sspra}.

In fact, the semiclassical expression (\ref{Gutztrace}) is consistent
with the semiclassical limit to the exact expression (\ref{modKreinNc}):
In Ref.\cite{hwg97} it was argued for the two-sphere case and in 
Ref.\cite{wreport} it was shown for any $N$-disk case that semiclassically
\beq
  \det M(k) \to \exp\left[-\sum_p\sum_{n=1}^\infty \frac{1}{n}\,
  \frac{{\rm e}^{ {\rm i} n k l_p -{\rm i}\nu_p\pi/2}}{\left|\det
  \left(\left[{\sf M}_p\right]^n-\openone\right)\right|^{1/2}}\right],
  \label{GutzVoros}
\eeq
where $l_p$, ${\sf M}_p$ and $\nu_p$ are the total geometrical length, 
the monodromy matrix and
the Maslov index of the $p$-th primitive periodic orbit, respectively.
The r.h.s.\ of (\ref{GutzVoros}) is the 
Gutzwiller-Voros zeta function~\cite{DasBuch}.
Note that for our scalar Dirichlet case there exists only one orbit, 
the bouncing-orbit for the two-sphere (two-disk) system with $l_p=2 L$, 
and that the Maslov index is simply
$\nu_p=4$  because of the two Dirichlet reflections 
(per repeat).~\footnote{
For the asymptotic 
case $k a_i \gg 1$ the scattering amplitude (\ref{Amplitude}) 
of the previous section
simplifies under  the Debye approximation
of the Bessel and Hankel function~\cite{Abramowitz} 
and the replacement of the angular momentum sum
by an integral:
\[
  A(ka_i) = \sum_{l=0}^{\infty} (-1)^l (2l+1)\frac{j_l(ka_i)}
 {h_l^{(1)}(ka_i)}
 \approx {\rm i} \frac{k a_i}{2} \exp(-{\rm i} 2 k a_i).
\]
This implies 
\[
  \det M^{{\rm o}_1{\rm o}_2} \!
  \approx 1-\frac{{\rm e}^{2{\rm i}kr}}{({\rm i} kr)^2}
  A(k a_1) A(k a_2)\\
  \approx 1-\frac{a_1 a_2}{4 r^2}\,{\rm e}^{{\rm i}2 k(r-a_1-a_2)},
\]
such that
\begin{eqnarray*}
{\cal N}_{\rm asym}^{{\rm o}_1 {\rm o}_2}&\approx& 
  -\frac{1}{\pi}{\rm Im}
  \ln\left[1-\frac{a_1 a_2}{4 r^2}\,{\rm e}^{{\rm i}2 k(r-a_1-a_2)}\right]\nn\\
 \\ &\approx&
  \frac{a_1 a_2}{4\pi r^2}\,\sin\left[2 k(r-a_1-a_2)\right]\,.
\end{eqnarray*}
The next term in the $1/kr$ expansion of the Hankel function 
(\ref{Hankel_asymp}) generates the correction
\bea
  {\cal N}_{\rm asym}^{{\rm o}_1 {\rm o}_2} &\approx&
  \frac{a_1 a_2}{4\pi r^2}\left(1+\frac{a_1}{r}+\frac{a_2}{r}\right)
  \sin\left[2 k(r-a_1-a_2)\right]\nn\\
  &\approx&  \frac{a_1 a_2}{4\pi r (r-a_1-a_2)}\sin\left[2 k(r-a_1-a_2)\right]
  \nn
\eea
which is the $n=1$ term of the Gutzwiller formula\,(\ref{Gutztrace}), 
consistent
with the asymptotic limit $L> a_i \gg 1/k$. This is  Eq.\,(13) 
of \cite{bw01} in the case $a_1=a_2$.}

If the semiclassical expression (\ref{Gutztrace}) is
inserted into the Casimir-energy  integral 
 (see \cite{bw01})
\bea
  {\cal E}_{\rm sc}&=&\half\hbar c \int_0^\infty {\rm d} k\,  k\, 
  \frac{{\rm d}\ }{{\rm d} k} {\cal N}_{\rm sc}(k)\nn\\
  &=& - \half \hbar c \int_0^\infty {\rm d} k\, {\cal N}_{\rm sc}(k)\,,
  \label{E_gen}
\eea
one gets for the scalar sphere-sphere case, after a Wick rotation
(as in the transition from (\ref{Ecint}) 
to  (\ref{EcWick})):~\footnote{The second 
line of Eq.\,(\ref{E_sc_o1o2}), after
the trivial integration over $k_4$, is identical with Eq.\,(3.20) of 
Ref.~\cite{sspra} if the latter is divided by a factor of two, as the
zero modes are weighted there with a factor of $1$ instead of $1/2$.}
\bea
  && {\cal E}_{\rm sc}^{{\rm o}_1 {\rm o}_2}(a_1,a_2,L) =   
  - \half \hbar c \int_0^\infty {\rm d} k\,
  {\cal N}^{{\rm o}_1 {\rm o}_2}_{\rm sc}(a_1,a_2,L,k)
\nn \\  &&
=  
-\half \hbar c \sum_{n=1}^{\infty}\frac{1}{n\, \pi}\, 
  \frac{\int_0^\infty {\rm d}k_4  \exp(- n 2 k_4 L)}
  { |\Lambda_+(a_1,a_2,L)^n+\Lambda_-(a_1,a_2,L)^n-2| } 
  \nn \\
  &&\approx- \frac{\hbar c}{16\pi}\,\frac{a_1 a_2}{L^2(a_1 +a_2+L)}
  \sum_{n=1}^{\infty}\frac{1}{n^4}\,
  \nn\\&&\ \text{}\times
  \left[1 -\frac{2}{3} \frac{L}{a_1\!+\!a_2}(n^2\!-\!1)
  -\frac{1}{3}L\frac{a_1^2+a_2^2}{a_1 a_2(a_1\!+\!a_2)}(n^2\!-\!1)\right]\nn\\
  &&\approx- \frac{\hbar c}{16\pi}\,\frac{a_1 a_2}{L^2(a_1+a_2+L)} 
  \left(\frac{\pi^4}{90}\right)\Biggl[ 1
 \nn \\  &&\ \text{} 
- \frac{2L}{a_1\!+\!a_2}\left(\frac{5}{\pi^2}\!-\!\frac{1}{3}\right)
  -\frac{L(a_1^2+a_2^2)}{a_1 a_2(a_1\!+\!a_2)}
  \left(\frac{5}{\pi^2}\!-\!\frac{1}{3}\right)
  \Biggr].\nn\\
\label{E_sc_o1o2}
\eea
Here we applied the following identity
\bea
  && \Lambda_+(a_1,a_2,L)^n+\Lambda_-(a_1,a_2,L)^n-2 \nn\\
  &&= 
  4n^2\Biggl[\frac{L}{a_1}+\frac{L}{a_2}+\frac{L^2}{a_1a_2}
  +\frac{2}{3}\frac{L^2}{a_1 a_2} (n^2-1)\nn\\
  &&\qquad\ \text{}
  +\frac{1}{3} \left(\frac{L^2}{a_1^2}+ \frac{L^2}{a_2^2}\right) 
(n^2-1)
  +{\cal O}\left(\frac{L^3}{a_i^3}\right)\Biggr],
  \label{Lambda_n}
\eea
which is exact for the case $n=1$, and used 
\[
\sum_{n=1}^\infty \frac{1}{n^4} =\frac{\pi^4}{90}\quad\text{and}\quad
\sum_{n=1}^\infty \frac{1}{n^2} =\frac{\pi^2}{6}\,.
\]

Particularly, in the case of two identical spheres of radius $a$ at 
center-to-center separation $R=L + 2a$ one gets a simplified expression
\bea
  &&{\cal E}_{\rm sc}^{{\rm o o}}(a,L) \equiv
  {\cal E}_{\rm sc}^{{\rm o}_1 {\rm o}_2}(a,a,L)=\nn\\
  && - \frac{\hbar c}{16\pi}\frac{a^2}{L^2(2a\!+\!L)} 
  \left(\frac{\pi^4}{90}\right)\!
  \left[1 
  \!-\! 2\left(\frac{5}{\pi^2}\!-\!\frac{1}{3}\right)\frac{ L}{a}
  \!+\!{\cal O}\!\left( \frac{L^2}{a^2}\right) 
  \right]\nn\\
  &&= \kappa \frac{\pi a^2}{L^2(2a\!+\!L)}\left[1 
  -2 \left(\frac{5}{\pi^2}\!-\!\frac{1}{3}\right)\frac{L}{a}
  +{\cal O}\left( \frac{L^2}{a^2}\right) 
  \right]\,.
 \label{E_ss_sc}
\eea 
It is remarkable that the leading term is exactly equal to the
plate-based prediction of the proximity-force approximation for two identical
spheres~\cite{sspra} (more details in Appendix~\ref{app:PFA}):
\bea
  \label{E_C_PFAoo}
  {\cal E}_{\rm plate\,PFA}^{{\rm o o}} 
  &=& \kappa\frac{\pi a^2}{L^2(2a+L)}\;.
\eea

As mentioned above, the sphere-plate system is a special case of the
two-sphere system:
\bea
  &&{\cal E}_{\rm sc}^{{\rm o} |}(a,L) \equiv \lim_{a_2\to\infty}
  {\cal E}_{\rm sc}^{{\rm o}_1 {\rm o}_2}(a,a_2,L)=\nn\\
  &&- \frac{\hbar c}{16\pi}\,\frac{a}{L^2} 
  \left(\frac{\pi^4}{90}\right)
  \left[1 
  -\left(\frac{5}{\pi^2}-\frac{1}{3}\right)\,\frac{L}{a}
  +{\cal O}\left( [L/a]^2\right) 
  \right]\nn\\
  && = \kappa \frac{\pi a}{L^2}\left[1 
  -\left(\frac{5}{\pi^2}-\frac{1}{3}\right)\,\frac{L}{a}
  +{\cal O}\left( [L/a]^2\right) 
  \right]\,.
  \label{E_sp_sc}
\eea
This expression agrees 
with Eq.\,(11) of Ref.~\cite{schroeder}.
Note that
\beq
  -{\cal E}_{\rm sc}^{{\rm o} |}(a,L) 
  < -\kappa \frac{\pi a}{L^2}  \,.
\eeq 
Moreover, the leading $\kappa \pi a/L^2$ behaviour of (\ref{E_sp_sc}) 
agrees with
the leading terms of the plate-based and sphere-based proximity-force
approximations for the Dirichlet sphere-plate problem, respectively
(see Refs.~\cite{gies1,jaffe,scardI})
\bea
  \label{E_C_PFAplate}
  {\cal E}_{\rm plate\,PFA}^{{\rm o}|} 
  &=& \kappa\frac{\pi a}{L^2}\,\frac{1}{1+L/a}\,,\\
  \label{E_C_PFAsphere}
  {\cal E}_{\rm sphere\,PFA}^{{\rm o}|} 
  &=& \kappa\frac{\pi a}{L^2}\,\Biggl\{1-3{\textstyle\frac{L}{a}}
  -6\left({\textstyle\frac{L}{a}}\right)^2   \nn \\
&& \mbox{}\times \Bigl[ 1 -\left(1+
  {\textstyle\frac{L}{a}}\right)
  \ln(1+a/L)\Bigr]\Biggr\}.
\eea
More details about the proximity-force approximation of the Dirichlet 
sphere-plate system can be found in Appendix~\ref{app:PFA}.


The 
sphere-plate result (\ref{E_sp_sc}) can also be derived
from the Dirichlet part of the identical-two-sphere result using
\bea
  {\cal N}_{\rm sc}^{\rm o|}
  &=& \left. {\cal N}_{\rm sc}^{\rm o o}\right|_{\rm D} 
  \nn\\
  &=&\frac{1}{\pi}{\rm Im}\sum_{n=1}^\infty 
  \frac{1}{n}\frac{\exp\left[{\rm i}n (r-2a)k\right]}
  {\left|\Lambda_+(a,r)^{n}+\Lambda_-(a,r)^{n}-2\right|}\,
  \label{N_sp_sc}
\eea
where (compare with Eqs.~(\ref{Gutztrace}) and (\ref{Lambda_pm}))
\bea
  \Lambda_\pm(a,r)&\equiv&\frac{r}{a}-1\pm 
  \sqrt{\left(\frac{r}{a}-1\right)^2-1}\nn\\
                &=&1 + 2\frac{L^{\rm o|}}{a}
  \pm \sqrt{\left(1+2\frac{L^{\rm o|}}{a}\right)^2-1 }\nn\\
                &=&\lim_{a_2\to\infty}\Lambda_{\pm}(a_1=a,a_2,L=L^{\rm o|})
\eea
under $r=2(L^{\rm o|}+a)$.
Note that
\bea
  \Lambda_\pm(a,r)^2 &=&\Lambda_{\pm}(a_1=a,a_2=a,L=L^{\rm oo})
\eea
for $r= L^{\rm oo}+2a$, such that for two identical spheres the
integrated density of states is semiclassically 
\beq
  {\cal N}_{\rm sc}^{\rm oo}
  =\frac{1}{\pi}{\rm Im}\sum_{n=1}^\infty 
  \frac{1}{n}\frac{\exp\left({\rm i}n 2(r-2a) k\right)}
  {\left|\Lambda_+(a,r)^{2n}+\Lambda_-(a,r)^{2n}-2\right|}\,.
\eeq

\section{Results}
\label{sec:results}
\begin{figure*}[h,t,b]

\psfrag{L/a}{\LARGE$L/a$}
\psfrag{-E/hbar c pi**3/1440 L**2}
{\LARGE${\cal E}^{\rm o|}_{\rm C}(a,L)\, /\,\frac{-\hbar c\pi^3 a}{1440 L^2}$}

\centerline{\includegraphics[width=12cm,angle=-90]{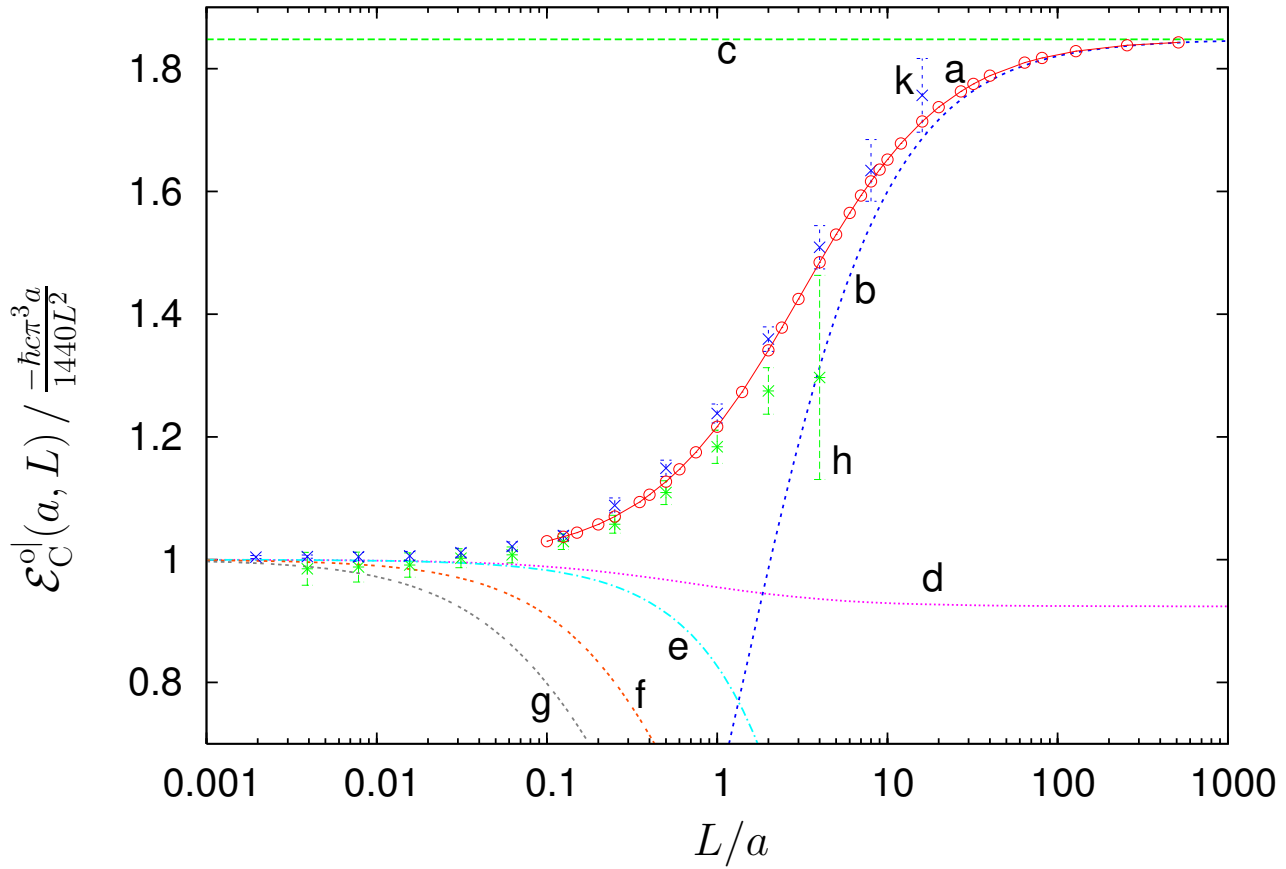}}

\caption{Predictions for the scalar Casimir energy 
${\cal E}_{\rm C}^{0|}(a,L)$ of the 
sphere-plate configuration with Dirichlet boundary conditions are shown 
in units of $\kappa \pi a/L^2$ 
as function
of the ratio $L/a$. The points and curves are explained in the text, see
Section \ref{sec:results}.
\label{fig:E_C-plots}}
\end{figure*}

The results and approximations, discussed in the previous sections for
the Casimir energy ${\cal E}_{\rm C}^{0|}(a,L)$ for the scalar Dirichlet 
sphere-plate case 
in units of  $\kappa \pi a/L^2=-\hbar c \pi^3 a/(1440 L^2)$ 
are shown
in Fig.\,\ref{fig:E_C-plots} 
as function
of the ratio $L/a$. 
This figure should be compared with
Fig.\,8 of the world-line approach of Ref.\,\cite{gies1} and with 
Fig.\,4 of the optical approach of Ref.\cite{scardI} which both 
only present
data for $L/a \leq 4$.
The circles (a) represent the numerically calculated exact expression 
(\ref{E_sp_exact}) for
the sphere-plate system between $L=0.1a$ and $L=512a$ (the line
is only there to guide the eye), 
the curve (b) shows the $s$-wave approximation
(\ref{E_sp_s-wave}), the line (c) represents the asymptotic limit 
1.847 of
(\ref{E_sp_asymp}), the curve (d) represents the numerically calculated 
(Wick-rotated) 
integral (\ref{E_gen}) over 
the semiclassical expression (\ref{N_sp_sc}) including {\em all} repeats,
and
the line (e) shows the analytical semiclassical formula
(\ref{E_sp_sc}) valid modulo ${\cal O}\bigl( (L/a)^2\bigr)$ corrections. 
The curve (f) is the result of the 
plate-based proximity-force approximation (\ref{E_C_PFAplate}), and the
curve (g)
represents the result of the sphere-based proximity-force approximation
(\ref{E_C_PFAsphere}). 

Our numerically calculated data (a) agree for $L/a\leq 1$ 
with the in Ref.\,\cite{gies1} published 
data of the world-line
approach, within  the quoted (statistical) error bars
which are already
sizable at $L/a=1$. We have included these world-line data, which cover 
the range from $L=a/256$ to $L=4a$, in 
Fig.\ref{fig:E_C-plots} and marked them by stars (h).   
Note that 
their central values are systematically on the low side in comparison to 
ours. 
All our data beyond $L=4a$ are 
predictions. In the meantime, after the first version of this paper
was submitted, there appeared in \cite{gies2005} 
new data in the world-line approach with improved systematics
which cover the extended range from $L= a/512 $ to $L=16a$ and which 
have smaller, but still sizable error bars for $L\geq a$ 
(see the crosses (k) in Fig.\,\ref{fig:E_C-plots}).
In the region of overlap, thus now also for the points $L/a=2, 4, 8, 16$,
the new world-line data  do nicely agree with ours when the quoted 
statistical error bars
are taken into account, although their central
values are now systematically higher. 
It should also be remarked that 
our smallest point $L=0.1a$ is  already affected
by a sizable truncation error in the integration. 
Of course this problem is a matter
of numerics and not a matter of principle.

Note that the $s$-wave approximation becomes a reasonable approximation to
the exact data from $L\ge 4a$  (it works very nicely for $L>15 a$ 
in agreement with the estimate
(\ref{lmax})) and, moreover, that
the exact data indeed converge to the 
predicted asymptotic value $180/\pi^4$ of Eq.\,(\ref{E_sp_asymp})
and do not show any Casimir-Polder $a^3/L^4$ scaling.
It is also interesting that the exact data, at least for the $L$ values
for which they could be calculated, are larger than 1 (in units
of $\kappa\pi a/L^2$), whereas the semiclassical approximation and the
proximity-force approximations are strictly less than 1 (in the same units).
It should be noted that (at least) the upper error ranges of 
the old world-line data of Ref.~\cite{gies1},
the complete new world-line data of 
Ref.~\cite{gies2005} (with the exception of the lower error ranges 
of the points $L/a=1/64,1/128$)
 and -- below $L\sim 0.1a$ -- the 
results of the improved
optical approach of Ref.~\cite{scardI} are 
larger than 1 as well.

The semiclassical calculation, 
starts out at a value 1 (in units of 
$\kappa \pi a/L^2$) for $L/a\to 0$,
which was predicted by the proximity-force approximation(s)~\cite{prox1,prox2}
and confirmed in, e.g., Refs.~\cite{sspra,ssprl,gies1,jaffe,scardI}.
Then, for intermediate values of $L/a$, however, the semiclassical results,
even though smaller than 1, are
superior to the results of the proximity-force approximation which become
ambiguous~\cite{gies1,jaffe,scardI}. 
For $L\gg a$, the contributions of the repeats of the
bouncing-orbit are strongly suppressed and the numerically calculated 
semiclassical expression  
converges to the one-bounce 
result $-\hbar c a/(16\pi L^2)=(90/\pi^4)\times \kappa \pi a/L^2$ 
which is smaller by a factor $1/2$ than the exact asymptotic answer
(\ref{E_sp_asymp}).

\section{Conclusions}
\label{sec:conclusions}
We presented here an exact calculation of the scalar Casimir energy for the
case of two spheres and a sphere and a plane. 
It is based
on a new Krein-type formula which directly expresses 
the geometry-dependent
part of the density of states by the inverse multi-scattering matrix of the
pertinent scattering problem. Thus the 
corresponding Casimir energy follows from
the energy-integral over the multi-scattering phase shift 
(the logarithm of the multi-scattering 
determinant). The calculation  is therefore  not
plagued by subtractions of the single-sphere contributions or by a 
removal 
of diverging ultra-violet contributions.  
The asymptotic limit (\ref{E_sp_asymp}),
the presented $s$-wave approximation (\ref{E_sp_s-wave}) and
all data with $L>4a$
are totally new results. 
Moreover, contrary to claims in the literature, the
Casimir-Polder scaling of the scalar Casimir effect is excluded by
our numerical {\em and} analytical calculations.

The two-sphere and sphere-plane cases are only two
examples, and the formalism presented can be easily extended to any number of
spheres and planes as well (or disks and lines in two-dimensions).  
We have exemplified the calculation of the scalar
Casimir energy only for the case of Dirichlet boundary conditions. One can
replace the Dirichlet with Neumann boundary conditions or with any other
conditions easily, or even replace the scatterers with arbitrary
nonoverlapping potentials/nonideal reflectors.

Aside from the exact results we have also discussed several approximation
schemes, the large separation limit, the semiclassical limit and the 
proximity-force approximation.  
The exact results (which are easy to calculate and
are definitely simpler to evaluate than in a path integral approach) should be
looked upon as test examples for other approximate methods. These results
already show that the proximity formula and the semiclassical/orbit approaches
are limited to small separations only, typically much 
smaller than the curvature
radii of the two surfaces.

One can make the argument that the dominant momenta of the fluctuating fields
contributing to the Casimir energy at a separation $L$ are of the order
$k\approx 1/L$ (see Eq.\,(\ref{lmax})) 
and thus for large separations only the $s$-wave scattering is
important. This is the main reason why the semiclassical approximation (which
is valid when many partial amplitudes contribute) fails at large separations.

Just the opposite is true at small separations, where semiclassics is pretty
good and so is 
the proximity formula, when the separations are smaller 
by one or, respectively, two orders of magnitude than
the curvature radii. The same type of analysis can be straightforwardly
extended to cylinders, or even objects with less symmetry (in which case the
corresponding individual T-matrices appearing 
in the inverse 
multi-scattering matrix (see Ref.\,\cite{Lloyd,Lloyd_smith,wreport})
become somewhat more complicated, due to
the loss of spherical symmetry).

\begin{acknowledgments}
A.W.\ thanks
Holger Gies and Antonello Scardicchio for useful discussions at QFEXT'05.
We are grateful to
Holger Gies
for supplying us
with the new  world-line data \cite{gies2005} prior to publication and for
helpful comments.  
Support from the Department of Energy under grant DE-FG03-97ER41014, from the
Polish Committee for Scientific Research (KBN) under Contract
No.~1~P03B~059~27 and by the Forschungszentrum J\"ulich under contract No.
41445400 (COSY-067) is gratefully acknowledged. 
\end{acknowledgments}

\appendix

\section{Ambiguity in the proximity-force approximation}
\label{app:PFA}
The proximity-force approximation (PFA) for the Casimir energy  
${\cal E}_{\rm C}$  of  two arbitrary smooth surfaces (with Dirichlet
boundary conditions for the scalar-field case)
is given by the  surface integral over the Casimir energy per area
which belongs to
an equivalent parallel-plate system  that locally follows the
two surfaces~\cite{prox1,prox2,gies1,jaffe,scardI}
, {\it i.e.}
\beq
  \label{PFA}
  {\cal E}_{\rm PFA} =\iint_{A} 
  {\rm d}\sigma\  \epsilon\left[z(\sigma)\right].
\eeq
Here $A$ is the area of one of the opposing surfaces which are locally
separated by the (surface-dependent) distance $z(\sigma)$ and
$\epsilon\left[z(\sigma)\right]$ is the corresponding 
Casimir energy per area. 
In general, the plate  segment  ${\rm d}\sigma$  is  tangential to only 
{\em one} of the surfaces and  therefore  the local 
distance vector ${\vec{z}}\,(\sigma)$ is perpendicular only to this 
surface and not to the other one.
Thus ${\cal E}_{\rm PFA}$ 
is not uniquely defined, since the area $A$ can be either one of the two
opposing surfaces (or even one ficticious surface somewhere inbetween). 
Particularly, for the case of a sphere of radius $a$ and a plate at 
shortest surface-to-surface separation $L$  we get the following expression for
the ``sphere-based PFA'' \cite{gies1,jaffe,scardI}, 
where the local distance vector $\vec z(\vec a)$
is  perpendicular to the sphere,  
\bea
  \label{PFAsphere}
  {\cal E}_{\rm sphere\,PFA}^{{\rm o}|} 
  &=&\kappa \iint_{\rm half-sphere}a^2 {\rm d} \Omega^{(2)}\, 
  \frac{1}{\left|\vec{z}(\vec a)\right|^3}\nn\\
  &=&\kappa a^2 {2\pi}\int_{0}^{\pi/2}{\rm d}\theta\,
  \sin(\theta)\,\frac{\left[\cos(\theta)\right]^3}
  {\left[ L+a -a\cos(\theta)\right]^3}\nn\\
  &=& \kappa\frac{\pi a}{L^2}\,\Biggl\{1-3{\textstyle\frac{L}{a}}
  -6\left({\textstyle\frac{L}{a}}\right)^2
  \nn \\&& 
\mbox{}\times\Bigl[1 -\left(1+
  {\textstyle\frac{L}{a}}\right)
  \ln(1+a/L)\Bigr]\Biggr\}\,.
\eea
The coefficient $\kappa$ is again the prefactor
$ 
 \kappa = -\frac{\hbar c \pi^2}{1440}
$
of the Casimir energy 
$
 {\cal E}^{||}(L)=-\frac{\hbar c\pi^2}{2\times 720}\,\frac{A}{L^3}
$
of the corresponding  scalar (Dirichlet) two-plate system.

On the other hand, 
the ``plate-based PFA'' \cite{gies1,jaffe,scardI} ({\em i.e.},
the local distance vector ${\vec{z}}\,(\sigma)$ is
perpendicular to the plate) follows from
\bea
  \label{PFAplate}
  {\cal E}_{\rm plate\,PFA}^{{\rm o}|} 
  &=&\kappa\iint_{x^2+y^2\leq a^2} 
  {\rm d}x\, {\rm d}y\,\frac{1}{\left|\vec{z}(x,y)\right|^3} \nn\\
  &=&\kappa \int_0^{2\pi}{\rm d}\phi\, \int_0^a{\rm d}\rho\, \rho\,
     \frac{1}{\left[L+a-\sqrt{a^2-\rho^2}\right]^3}\nn\\
  &=& \kappa\frac{\pi a}{L^2}\,\frac{1}{1+L/a}\,.
\eea

Note that 
\beq
   -{\cal E}_{\rm sphere\,PFA}^{{\rm o}|}
  <  -{\cal E}_{\rm plate\,PFA}^{{\rm o}|}
  <-\kappa \frac{\pi a}{L^2}\,.
\eeq

Finally, the PFA for two spheres of a common radius $a$ and shortest 
surface-to-surface separation $L$ can be derived from the plate-based PFA
(\ref{PFAplate}) of
the sphere-plate case, with the fictitious 
plate on the vertical symmetry plane as
in Fig.\,\ref{fig:twosphere}, as follows:
\bea
  \label{PFAoo}
  {\cal E}_{\rm plate\,PFA}^{{\rm o o}} 
  &=&\kappa\iint_{x^2+y^2\leq a^2} 
  {\rm d}x\, {\rm d}y\,\frac{1}{\left|2\, \vec{z}(x,y)\right|^3} \nn\\
  &=&\kappa \pi\int_0^{2\pi}\!\!{\rm d}\phi\, \int_0^a
\!\!{\rm d}\rho\, \rho\,
     \frac{1}{8\left[\frac{L}{2}+a-\sqrt{a^2-\rho^2}\right]^3}\nn\\
  &=& \kappa\frac{\pi a}{8 \left(\frac{L}{2}\right)^2}\,\frac{1}{1+L/2a}
  = \kappa\frac{\pi a^2}{L^2(2a+L)}\;.
\eea

\section{Comparison with the two-dimensional two-disk and disk-line
systems}
\label{app:twodim}

The two-dimensional analog of the $N$-sphere matrix 
in three-dimensions (\ref{M_N-sphere}) 
is \cite{wreport,wh98}
\bea
&& {\bf M}^{j j'}_{m m'} = \delta^{jj'}\delta_{mm'} +(1- \delta^{jj'})
                  \frac{ a_j   \Jb{m }{ka_j} }
                       { a_{j'}\Hone{{m'}}{k a_{j'}}  }\nn\\
&&\quad\text{}\times                  (-1)^{m'}
                  {\rm e}^{{\rm i} ( m \alpha_{j' j} - m' \alpha_{j j'} ) }
                  \Hone{m-m'}{k r_{jj'}}  ,
 \label{M_N-disk}
\eea
where $j,j'=1,2,\dots,N$ are the labels of the $N$ disks. The integers 
$m,m'$ with $-\infty <m,m'<\infty$ are the angular momentum quantum numbers in
two-dimensions, $a_j$ and $r_{jj`}$ are, as usual, the radius of disk $j$ and
the distance between the centers of disk $j$ and $j'$, respectively. 
$\Jb{l}{k r}$ and
$\Hone{l}{k r}$ are the ordinary Bessel and Hankel functions of first
kind, and $\alpha_{jj'}$ is the angle of the  of the ray 
from the origin of
disk $j$ to the one of disk $j'$ as
measured in the local coordinate system of disk $j$.

The two-dimensional analog of the two-sphere KKR-type matrix (\ref{M_2-sphere})
is~\cite{aw_chaos,aw_nucl}
\beq
   \mat{{M}^{11}}{{M}^{12}} 
       {{M}^{21}}{{M}^{22}}_{m, m'}
=  \mat{ \delta_{m m'} }{ A_{mm'}^{12} }
  { A_{mm'}^{21}   }{ \delta_{m m'} }
 \label{M_2-disk}
\eeq
with
\bea
  &&  A_{mm'}^{jj'}= (-1)^{m}\, \frac{ a_j   \Jb{m }{ka_j} }
                       { a_{j'}\Hone{m'}{k a_{j'}}}   \Hone{m-m'}{k r}\,,
  \label{A-2dim}
\eea
where $r=r_{12}=r_{21}$.
The general two-disk system is characterized by a $C_{2v}$ symmetry. If the
two disks have a common radius $a$, the global domain of this system
is separated by the vertical symmetry {\em axis} (instead of
plane) into two half-domains. The corresponding symmetry group is now 
$C_{2 h}$ and the KKR-matrix splits into two KKR matrices valid for
the half-domains, one corresponding to Neumann (N) boundary conditions of
the scattering wave functions  on the
vertical symmetry axis, the other, with the additional minus sign, 
corresponding
to the Dirichlet (D) case:
\bea
  \left. M^{{\rm oo}}_{mm'}\right|_{\rm N} &=& \delta_{mm'} 
  + A_{mm'}\,,
  \label{M_Neumann_2}
  \\
  \label{M_Dirichlet_2}
   \left. M^{{\rm oo}}_{mm'}\right|_{\rm D} 
   &=& \delta_{mm'} - A_{mm'}\,,
\eea
where $A_{mm'}=A^{12}_{mm'}|_{a_1\!=\!a_2\!=\!a} 
= A^{21}_{mm'}|_{a_1\!=\!a_2\!=\!a}$.

The two-dimensional KKR-matrix in the large-distance limit reads
\beq
  [ M (k)]^{jj'}  \approx
  \delta^{jj'} - (1-\delta^{jj'}) f_j^{{\rm 2D}}(\veps)
  \frac{ \exp ({\rm i} kr_{jj'})}{\sqrt{r_{jj'}}}   \label{eq:kkr0_2}
\eeq
instead of (\ref{eq:kkr0}).
Here $f_j^{{\rm 2D}}(\veps)$ is the $s$-wave scattering amplitude in
two dimensions.

Since the asymptotic expression of the ordinary Hankel function 
reads~\cite{Abramowitz}
\bea
  \Hone{m}{kr}&\sim& \sqrt{\frac{2}{\pi kr}}\exp\left[{\rm i}\left(
 kr -m\frac{\pi}{2}-\frac{\pi}{4}\right)\right]
\nn\\
&\sim& (-{\rm i})^m \Hone{0}{kr}\,,
  \label{Hankel_asymp_ord}
\eea
Eqs.~(\ref{M_Neumann_2}-\ref{M_Dirichlet_2}) become asymptotically 
\bea
   M^{{\rm oo}}_{m_1 m_2} (k, a, r) 
   &\sim& 
\delta _{m_1 m_2} \pm 
(-1)^{m_1} 
  \frac{\Jb{m_1}{ka}}{\Hone{m_2}{ka}}
\,\nn\\
 &&\qquad\text{}\times {\rm i}^{m_2-m_1}\,\Hone{0}{kr}
 \label{M_asymp_res_2}
\eea
instead of (\ref{M_asymp_res}). This expression is separable in 
$m_1$ and $m_2$.
Therefore, the corresponding 
determinant is asymptotically given by
\bea
  &&\det M^{\rm oo}(k ,a,r)\nn\\
  &&\sim 1 - 
  \left[\Hone{0}{kr}\right]^2\left [ \sum _{m=-\infty}^\infty (-1)^m
  \frac{\Jb{m}{ka}}{\Hone{m}{ka}}
  \right ]^2\nn\\
  &&\equiv 1 -  \left[\Hone{0}{kr}\right]^2\left[
A(ka)\right]^2.
  \label{detM_asymp_glob_2}
\eea
for the identical two-disk case and by
\bea
  &&\det M^{\rm oo}(k ,a,r)\nn\\
  &&\sim 1 - \Hone{0}{kr}
  \left [ \sum _{m=-\infty}^\infty (-1)^m
  \frac{\Jb{m}{ka}}{\Hone{m}{ka}}
  \right ]\nn\\
  &&\equiv 1 - \Hone{0}{kr}\, \left[
A(ka)\right]
  \label{detM_asymp_dir_2}
\eea
for the disk-line case. Note that the asymptotic 
relation\,(\ref{Hankel_asymp}) is actually exact for   $\hone{0}{kr}$ and
therefore holds also for small $k$-values.
For the ordinary Hankel function  $\Hone{0}{kr}$, however, the 
corresponding formula (\ref{Hankel_asymp_ord}) is only an asymptotic relation.
Since the Casimir energy receives contributions from  all
$k < 10/L$, it severely worsens the $s$-wave result if 
$\Hone{0}{kr}$ were replaced by its asymptotic form
from (\ref{Hankel_asymp_ord}).
Here
\beq
  A(ka)\equiv  \sum _{m=-\infty}^\infty (-1)^m 
  \frac{\Jb{m}{ka}}{\Hone{m}{ka}}
  \label{Amplitude_2}
\eeq
is the multipole expansion of the scattering amplitude in two dimensions. 
For $ka\ll 1$, the dominant effect comes from the $m=0$ ($s$-wave)
contribution and the $|m|>0$ terms can be neglected. The scattering amplitude
becomes~\cite{rww96}
\bea
  A(ka) &=&  \frac{\Jb{0}{ka}}{\Hone{0}{ka}} 
  +{\cal O}\left((ka)^2\right)\nn\\ 
  &=&   
  \frac{1}{1\!+\!{\rm i}\frac{2}{\pi}
\left[\ln\left(\frac{ka}{2}\right)\!+\!\gamma_E\right]} 
 +{\cal O}\!\left((ka)^2\right) ,
\eea
where $\gamma_E=0.577\cdots$ is Euler's constant.
For the asymptotic case $kr> k a_i\gg 1$, one finds 
\[ A(ka_i) \equiv  \sum _{m=-\infty}^\infty (-1)^m 
  \frac{\Jb{m}{ka_i}}{\Hone{m}{ka_i}}
\approx \frac{\sqrt{{\rm i}\pi  k a_i} }{2}\e^{-2{\rm i} k a_i}
\]
for the scattering amplitude,
such that
\bea
  \det M^{{\rm o}_1{\rm o}_2} 
  &\approx& 1-\frac{2}{\pi}\frac{\e^{2{\rm i}kr}}{{\rm i} kr}
  A(k a_1) A(k a_2)\nn \\
  &\approx& 
1-\frac{\sqrt{a_1 a_2}}{2 r}\,{\rm e}^{{\rm i}2 k(r-a_1-a_2)}\,.
\eea
In fact, the two-dimensional analogs of the semiclassical expression
(\ref{E_ss_sc})
for the identical two-sphere case and (\ref{E_sp_sc}) for 
the sphere-plate case are given by
\bea
{\cal E}^{{\rm oo}}_{{\rm sc}}
&\approx&-\frac{\hbar c}{8\pi}\,\frac{a}{L^{3/2}\sqrt{2a +L}}
\sum_{n=1}^\infty \frac{1}{n^3}\nn\\
&=&-\frac{\hbar c}{16\pi}\zeta(3)\frac{\sqrt{2}}{L}\sqrt{\frac{a}
{L(1+L/2a)}} 
\eea
for the two-disk case and
\bea
{\cal E}^{{\rm o|}}_{{\rm sc}}
&\approx&-\frac{\hbar c}{2}\,\sum_{n=1}^\infty \frac{1}{\pi
  n} \frac{1}{2n L} \frac{1}{2n}\sqrt{\frac{a}{L}}\nn\\
&=& -\frac{\hbar c\zeta(3)}{16\pi} \frac{2}{L} \frac{\sqrt{a}}{\sqrt{L}}
= \kappa_{\rm 2D}\, 2 \frac{\sqrt{a}}{L\sqrt{L}}
\label{E_dl_sc}
\eea
for the disk-line case, where 
\beq
 \kappa_{\rm 2D}\equiv -\frac{\hbar c}{16\pi}\zeta(3) \ \   \mbox{and} 
\ \ 
 \zeta(3)\equiv \sum_{n=1}^{\infty} \frac{1}{n^3} \approx 1.20205.
\eeq
The corresponding proximity-force approximation reads in the
line-based scenario
\bea
  {\cal E}_{\rm line\,PFA}^{{\rm o}|} 
  &=& \kappa_{\rm 2D}  \int_{-a}^{a} {\rm d}x\,\frac{1}{\left(L + a
  -\sqrt{a^2-x^2}\right)^2}\nn\\
  &=& 2 a^2\kappa_{\rm 2D} 
\frac{2\arctan\left(\frac{L+2a}{\sqrt{L(L+a)}}\right) 
+\frac{\sqrt{L(L+2a)}}{a}}
{L (L+2a)\sqrt{L(L+2a)}} \nn\\
&\approx&\kappa_{\rm 2D}\frac{\pi}{\sqrt{2}}\frac{\sqrt{a}}{L\sqrt{L}}
\approx \kappa_{\rm 2D}\, 2.22144\frac{\sqrt{a}}{L\sqrt{L}}\,.
  \label{E_C_PFAline}
\eea
The proximity-force approximation for the disk-line system in the
disk-based scenario is given by
\bea
  {\cal E}_{\rm disk\,PFA}^{{\rm o}|} &=& \kappa_{\rm 2D}\, 
a \int_{-\pi/2}^{\pi/2}
{\rm d}\phi \frac{\left[\cos \phi\right]^2}{\left(L+a-a\cos\phi\right)^2}\nn\\
 &\approx&   \kappa_{\rm 2D}\frac{\pi}{\sqrt{2}}\frac{\sqrt{a}}{L\sqrt{L}}
\approx \kappa_{\rm 2D}\, 2.22144\frac{\sqrt{a}}{L\sqrt{L}}\nn\\
 \label{E_C_PFAdisk}
\eea
as well.
Note that in the limit $L/a\to 0$ the two-dimensional semiclassical
approximation (\ref{E_dl_sc}) and the proximity-force approximations
(\ref{E_C_PFAline}) and (\ref{E_C_PFAdisk}) do approximately agree, but
are not identical. Furthermore, the exact result, in the range where it can
be calculated, {\it i.e.}, for $L > 0.1 a$, does
not scale as $\sqrt{a/L}/L$, but rather as
 $(a/L)^{1/6}/L$. The $s$-wave result is a good approximation to the
exact result for $L> 10a$, if $\Hone{0}{kr}$ is not replaced by its
asymptotic form (\ref{Hankel_asymp_ord}).


\end{document}